\documentclass[12pt]{article}
\usepackage{amssymb,amsmath}
\usepackage[noblocks]{authblk}
\usepackage[top=0.75in, bottom=0.75in, left=0.75in, right=0.75in, dvips]{geometry}
\usepackage{caption}
\begin{document}
\textwidth 10.0in
\textheight 9.0in
\topmargin -0.60in
\title{A Massive Non-Abelian Vector Model}
\author[1,3] {F.A. Chishtie}
\author[1,2]{D.G.C. McKeon}
\affil[1] {Department of Applied Mathematics, The
University of Western Ontario, London, ON N6A 5B7, Canada}
\affil[2] {Department of Mathematics and
Computer Science, Algoma University, Sault St.Marie, ON P6A
2G4, Canada}
\affil[3] {Department of Space Science, Institute of Space Technology, Islamabad 44000, Pakistan}
\maketitle

\maketitle

\begin{abstract}
The introduction of a Lagrange multiplier field to ensure that the classical equations of motion are satisfied serves to restrict radiative corrections in a model to being only one loop.  The consequences of this for a massive non-Abelian vector model are considered.
\end{abstract}
\vspace{.3cm}\\

The elusiveness of the Higgs Boson has led to reconsideration of various ways of endowing a non-Abelian vector field with a mass.  For a $U(1)$ vector field, Stueckelberg has shown that such a mass can be inserted ``by hand'' without compromising either unitarity or renormalizability [1].  Indeed, the $U(1)$ sector of the Standard Model may have such a mass, which  makes the masslessness of the photon somewhat mysterious [2,3].

The Lagrangian for a massive $SU(N)$ vector field $A^a_{\mu}$  ($a = 1 ... N$, where $N$ is the dimension of the gauge group),
\begin{equation}
\mathcal{L}_I(A) = -\frac{1}{4}  F_{\mu\nu}^a F^{a\mu\nu} - \frac{m^2}{2} A_\mu^a A^{a\mu}
\end{equation}
\begin{equation}
\hspace{-4cm}\left([D_\mu , D_\nu]^{ab} = c^{apb} F_{\mu\nu ,}^p \;, D_\mu^{ab} = \partial_\mu \delta^{ab} + c^{apb} A_\mu^p \;, \eta^{\mu\nu} = \mathrm{diag} (-,+,+,+)\right)\nonumber
\end{equation}
has been investigated with the hope that the symmetry present would be sufficient to ensure that this model for the vector $A_\mu^a$ is both unitary and renormalizable for $m^2 \neq 0$ even if the group were not $U(1)$ [4-23].  It has been shown that tree level unitarity is not upheld on account of the longitudinal polarization of $A_\mu^a$ [20-22] and that renormalizability is lost beyond one loop order [7, 12].

The equation of motion for $A_\mu^a$ is
\begin{equation}
D_\mu^{ab} (A) F^{b\mu\nu} - m^2 A^{a\nu} = 0;
\end{equation}
we can ensure that $A^{a\mu}$ satisfies this equation of motion by supplementing $\mathcal{L}_I$ with
\begin{equation}
\mathcal{L}_{II}(A,B) = B_\nu^a \left(D_\mu^{ab} F^{b\mu\nu} - m^2 A^{a\nu}\right)
\end{equation}
where $B_\nu^a$ is a ``Lagrange multiplier'' field. The Lagrangian $\mathcal{L} = \mathcal{L}_I + \mathcal{L}_{II}$ has been first investigated when $m^2 = 0$ in [24] and also later in [25].  In [24], it has been shown that perturbative radiative effects vanish beyond one loop order and that consequently the model can be considered to be ``solvable''.  We wish to now extend these considerations to the case $m^2 \neq 0$.

When $m^2 \neq 0$, no local gauge symmetry is present and so the generating functional is simply
\begin{equation}
Z[J_{\mu }^a, K_\mu^a ] = \int DA_\mu^a\, DB_\mu^a \exp \,i \int d^4x \left[ \mathcal{L} + J_\mu^a A^{a\mu} + K_\mu^a B^{a\mu}\right].
\end{equation}
The terms in the action that are bilinear in the fields are
\begin{equation}
\frac{1}{2} \left( A_{\mu }, B_\mu^a\right)
\left( \begin{array}{cc}
a^{\mu\nu} & a^{\mu\nu} \\
a^{\mu\nu} & 0
\end{array}
\right)
\left( \begin{array}{c}
A_\nu^a \\
B_\nu^a
\end{array}
\right)
\end{equation}
where $a^{\mu\nu} = (\partial^2 - m^2)\eta^{\mu\nu} - \partial^\mu\partial^\nu$. The inverse of the operator $\mathbf{M}$ in eq. (5) is
\begin{equation}
\mathbf{M}^{-1} =
\left( \begin{array}{cc}
0 & (a^{-1})_{\mu\nu} \\
(a_{\mu\nu}^{-1}) & -(a^{-1})_{\mu\nu}
\end{array}
\right)
\end{equation}
where
\begin{equation}
(a_{\mu\nu}^{-1}) = \frac{\eta_{\mu\nu} - \partial_\mu \partial_\nu/m^2}{\partial^2 - m^2} .
\end{equation}
From eq. (6), we see that there is a propagator $<BB>$ for the field $B_\mu^a$ as well as mixed propagators $<AB>$ and $<BA>$, but no $<AA>$ propagator for the field $A_\mu^a$.  This fact, combined having only the vertices $<AAA>$, $<AAAA>$, $<BAA>$, and $<BAAA>$ (ie, no vertex involves more than one external $B_\mu^a$ field), leads to the disappearance of all loop diagrams beyond one loop order, as in the $m^2 = 0$ case [24].  The one loop diagrams receive no contribution from the propagator $<BB>$ or from the vertices $<AAA>$, $<AAAA>$.

To see more directly how diagrams beyond one loop order cannot contribute to $Z$, we first perform the path integral over $B_\mu^a$ in eq. (4) to give
\begin{equation}
Z[J_{\mu }^a, K_\mu^a ] = \int DA_\mu^a\,\delta \left( \frac{\delta\mathcal{L}_I(A)}{\delta A_\lambda^p} + K^{a\lambda}\right)
 \exp \,i \int d^4x \left( \mathcal{L_I}(A) + J_\mu^a A^{a\mu}\right).
\end{equation}

This result follows from the fact that $B^a_{\mu}$ acts as a Lagrange multiplier field; we are using the functional analogue of $\int dx dy \ e^{i(h(x)+yf(x))}=\int dx \delta (f(x)) e^{ih(x)}$.
The standard result
\begin{equation}
\int_{-\infty}^\infty dx\; \delta(f(x)) g(x) = \sum_i g(a_i)/f^\prime(a_i)\quad(f(a_i) = 0)
\end{equation}
can next be used to evaluate the path integral over $A_\mu^a$ in eq. (8).  We obtain
\begin{equation}
Z[J_{\mu }^a, K_\mu^a] = \sum_i
 \left( \det \frac{\delta^2\mathcal{L}_I(A)}{\delta A_\lambda^p\delta A_\sigma^q}\right)^{-1}   \exp \,i \int d^4x \left( \mathcal{L}_I(A) + J_\mu^a A^{a\mu}\right)
\end{equation}
where the sum in eq. (10) is over those configurations that satisfy the equation of motion
\begin{equation}
\frac{\delta\mathcal{L}_I}{\delta A_\lambda^p} + K^{p\lambda} = D_\rho^{pq} F^{q\rho \lambda} - m^2 A^{p\lambda} + K^{p\lambda} = 0.
\end{equation}

We can relate the result of eq. (10) with what is obtained by quantizing the action involving only the Lagrangian $\mathcal{L}_I$ of eq. (1). This will show directly that as a consequence of having supplemented $\mathcal{L}_I$ with $\mathcal{L}_{II}$, we have a model that has no radiative corrections beyond one loop order. If we consider
\begin{equation}
\overline{Z}[K_\mu^a] = \int DA_\mu^a \exp i\int d^4x \left[\mathcal{L}_I(A) + K_\mu^a A^{a\mu}\right]
\end{equation}
and expand [28-29]
\begin{equation}
A_\mu^a = V_\mu^a + Q_\mu^a
\end{equation}
where $V_\mu^a$ is a solution to the classical equation
\begin{equation}
\frac{\delta \mathcal{L}_I}{\delta A_\mu^a} + K_\mu^a = 0
\end{equation}
and $Q_\mu^a$ is a fluctuation about $V_\mu^a$, then working to the term quadratic in $Q_\mu^a$ in the action we obtain
\begin{equation}
\overline{Z}[J_\mu^a,V_\mu^a] \approx \int DQ_\mu^a \exp i \int d^4x \left[ \mathcal{L}_I(V) +
 \frac{1}{2} Q_\mu^a \frac{\delta^2\mathcal{L}_I (V)}{\delta Q_\mu^a \delta Q_\nu^b} Q_\nu^b + K_\mu^a V^{a\mu}\right].
\end{equation}
The functional integral over $Q_\mu^a$ can be evaluated in eq. (15) to yield
\begin{equation}
\approx \left( \det \frac{\delta^2\mathcal{L}_I (V)}{\delta Q_\mu^a \delta Q_\nu^b}\right)^{-1/2} \exp i \int d^4x [\mathcal{L}_I(V) + K^a_\mu V^{a\mu}].
\end{equation}
It is known that the exponential in eq. (16) is a consequence of tree level diagrams while the functional determinant is a consequence of the one loop diagrams in the presence of a background field $V_\mu^a$ [26, 27].

Eqs. (10) and (16) differ in two respects. First of all, one must set $J_\mu^a = K_\mu^a$ in eq. (10).  Secondly, as the connected Green's functions are generated by $W = -i \ln Z$, it follows that the connected one loop Green's functions that follow from eq. (10) differ from those that follow from eq. (16) by a factor of $1/2$.

It has been argued that tree level amplitudes that follow from eq. (12) do not satisfy unitarity [20]. However, since explicit integration over the field $B^a_{\nu}$ in eq. (4) leads to eq. (10) with all contributions from tree diagrams being contained in the exponential occurring in eq. (10), we have a heuristic demonstration of why it is plausible that the generating functional of eq. (4) is consistent with unitarity. The configurations of $A^a_{\mu}$ that appear in eq. (10) are all solutions of the classical equations of motion; they incorporate the sum of all tree diagrams considered in ref. [20] and hence it is apparent that upon summing all tree diagrams, the unitarity violating portion of the tree amplitudes considered in ref. [20] will cancel. 

The renormalizability of the model described solely by the Lagrangian $\mathcal{L}_I(A)$ has been examined in refs. [4-23].  By comparing eqs. (10) and (16), we see that the connected Green's functions at one loop order (whose generating functional is given by $W = -i\ln Z$ [28, 29]) for the models defined by $\mathcal{L}_I$ and $\mathcal{L}_I + \mathcal{L}_{II}$ differ by a factor of $1/2$.  Consequently, if the model defined by $\mathcal{L}_I$ is renormalizable at one loop order, then the model defined by $\mathcal{L}_I + \mathcal{L}_{II}$ is completely renormalizable.

Any discussion of renormalizability based on a direct analysis of Feynman diagrams which follow immediately from $\mathcal{L}_I$ is complicated by the term $(k_\mu k_\nu/m^2)/(k^2 + m^2)$ appearing in the vector propagator (of eq. (7)).  This is because the usual power counting arguments normally used to establish renormalizability breakdown. One might try to employ operator regularization to treat the massive field as in ref. [17], as with this method of regularization, no explicit divergences ever arise. However, if we were to work at one loop order in this model, operator regularization forces one to use the result, 
\begin{equation}
\exp -i[k^2 L_{\mu\nu} + m^2 \eta_{\mu\nu}]t = e^{-im^2t}\left[L_{\mu\nu} e^{-ik^2t} + T_{\mu\nu}\right]
\end{equation}
($T_{\mu\nu} \equiv \eta_{\mu\nu}-k_\mu k_\nu / k^2 \equiv \eta_{\mu\nu} - L_{\mu\nu}$).  The term in eq. (17) proportional to $T_{\mu\nu}$ necessarily results in ill-defined loop momentum integrals, and hence use of operator regularization in conjunction with the model of eq. (1)  is problematic.

This difficulty can be overcome [7, 12] by adapting the Faddeev-Popov approach to quantizing the massless Yang-Mills theory [30].  By inserting in turn the constant factors
\begin{subequations}
\begin{align}
\mathrm{const} &= \int D\Omega \delta(\partial_\mu A^{\Omega \mu} - k)\Delta_{FP} \\
\mathrm{const} &= \int dk \,e^{\frac{-i}{2\alpha}\int dx\, k^2}
\end{align}
\end{subequations}
into the generating functional of eq. (12) ($A_\mu = T^a A_\mu^a,\, A_\mu^\Omega = \Omega (A_\mu + \partial_\mu)\Omega^{-1},\, \Omega = e^{i\phi}, \Delta_{FP} = \det (\partial^\mu(\partial_\mu\delta^{ab} + c^{apb} A_\mu^p))$) and performing the transformation $A_\mu^\Omega \rightarrow A_\mu$ the generating functional of eq. (12) becomes
\begin{align}
\overline{Z}[K_\mu^a] &= \int D\Omega \int DA_\mu^a \Delta_{FP} \exp i \int dx \left[ - \frac{1}{4} F_{\mu\nu}^a F^{a\mu\nu} - \frac{1}{2\alpha} (\partial \cdot  A^a)^2 \right. \\
 &  \hspace{2cm}  \left. - \frac{m^2}{2} Tr (A_\mu + \Omega \partial_\mu \Omega^{-1})^2 + K_\mu^a A^{a\mu}\right] . \nonumber
\end{align}

If only the lowest order contribution in $\phi$ is retained in eq. (19), then the integral over $\phi$ can be evaluated and the one loop generating functional is found to be
\begin{align}
\overline{Z}[K_\mu^a] &\approx \int DA_\mu^a \Delta_{FP}^{1/2} \exp i \int d^4x \left[ - \frac{1}{4} F_{\mu\nu}^a F^{a\mu\nu} - \frac{1}{2\alpha} (\partial \cdot  A^a)^2 \right. \\
 &  \hspace{2cm}  \left. - \frac{m^2}{2} A_\mu^a A^{a\mu} + K_\mu^a A^{a\mu}\right] . \nonumber
\end{align}
Any (one loop) divergences arising from this functional integral can be renormalized [7, 12].  (Indeed, it is shown in ref. [30] that eq. (20) also arises in the Yang-Mills-Higgs model at one loop order if the Higgs field were to be integrated out.)  As a result, we conclude that the model defined by $\mathcal{L}_I + \mathcal{L}_{II}$ is renormalizable.

In refs. [15-17] the Kunimasa-Goto [4] form of the measure Yang-Mills model is examined and it is purportedly shown that even at one loop order the model is not renormalizable.  However, this analysis is likely deficient as the implementation of background field quantization in these papers for the ``Stueckelberg'' field is not consistent with the analysis of this technique provided in refs. [28, 29]; consequently it is not clear if in fact it is one loop effects that are being considered in refs. [15-17].

A canonical analysis of the system described by $\mathcal{L}$ is straightforward.  The momenta that follow from
\begin{equation}
\pi^{a\mu} = \partial \mathcal{L}/\partial(\partial_0 A_\mu^a) \quad \tau^{a\mu} =  \partial \mathcal{L}/\partial(\partial_0 B_\mu^a) \nonumber
\end{equation}
are
\[ \pi^{a0} = \tau^{a0} = 0 \eqno(21a,b) \]
\[ \pi^{ai} = F_{0i}^a - D_i^{ab} b^b + \partial_0 B_i^a + c^{abc} a^b B_i^c \eqno(21c) \]
\[\tau^{ai} = F_{0i}^a = \partial_0 A_i^a - D_i^{ab} a^b \eqno(21d) \]
\hspace{2cm}$(a^a \equiv A_0^a, \,\, b^a \equiv B_0^a)$\\
which lead to the canonical Hamiltonian
\[ \mathcal{H} = - \frac{1}{2} (\pi^{ai} - \tau^{ai})^2 + \frac{1}{2} \pi^{ai}\pi^{ai} + \frac{1}{4} F_{ij}^a F_{ij}^a + (D_i^{ab} B_j^b)(F_{ij}^a)\eqno(22) \]
\[\hspace{-1cm} -a^a (D_i^{ab} \pi^{bi} + c^{abc} B_i^b \tau^{ai}) - b^a D_i^{ab} \tau^{bi} \nonumber \]
\[ \hspace{3cm}+ \frac{m^2}{2} \left[ (A^{ai} + B^{ai})^2 - B^{ai}B^{ai} - (a^a + b^a)^2 + b^ab^a\right].\nonumber \]
The primary constraints of eqs. (21a,b) lead to the secondary constraints
\[ D_i^{ab} \pi^{bi} + c^{abc} \tau^{ai} + m^2 (a^a + b^a) = 0 \eqno(23a) \]
\[ D_i^{ab} \tau^{bi} + m^2 a^a = 0; \eqno(23b) \]
all constraints are second class provided $m^2 \neq 0$ [32].  This Hamiltonian is not positive definite.

Together, eqs. (21 a, b; 23 a, b) constitute a set of $4N$ second class constraints in phase space of the $16N$ fields present ($A^a_{\mu}$, $B^a_{\mu}$ and their conjugate momenta). This leaves $12N$ degrees of freedom in phase space: $6N$ being the three polarizations of the massive vector and their conjugate momenta plus $6N$ similar degrees of freedom associated with the Lagrange multiplier field. Those degrees of freedom associated with the Lagrange multiplier field are integrated out when arriving at the expression of eq. (10). 

In order to incorporate a Fermion $\psi$, we supplement $\mathcal{L}_I$ of eq. (1) with
\[\mathcal{L}_I^\prime = \overline{\psi} (/\!\!\!p - /\!\!\!\!A^a T^a - \kappa)\psi \eqno(24a) \]
and $\mathcal{L}_{II}$ of eq. (3) with
\[\mathcal{L}_{II}^\prime = -\overline{\psi} /\!\!\!\!B^a T^a \psi + \overline{\eta} (/\!\!\!p - /\!\!\!\!A^a T^a - \kappa)\psi +
\overline{\psi} (/\!\!\!p - /\!\!\!\!A^a T^a - \kappa)\eta \eqno(24b) \]
where $\overline{\eta}$ and $\eta$ are, like $B_\mu^a$, Lagrange multiplier fields used to ensure that the equations of motion that follow from $\mathcal{L}_I^\prime$ are satisfied.  Again, no effects beyond one loop order arise when considering $\mathcal{L}_I + \mathcal{L}_I^\prime + \mathcal{L}_{II} + \mathcal{L}_{II}^\prime$.

To illustrate this, let us consider the Abelian limit of eqs. (1,3,24) so that
\[ \mathcal{L} = \frac{1}{4}\left( \partial_\mu A_\nu - \partial_\nu A_\mu\right)^2 - \frac{m^2}{2} A_\mu^2 + B_\nu \left[ \partial_\mu
\left(\partial^\mu A^\nu - \partial^\nu A^\mu\right) - m^2 A^\nu\right] \nonumber \]
\[ + \overline{\psi} (/\!\!\!p - /\!\!\!\!A - \kappa )\psi -
 \overline{\psi} \,/\!\!\!\!B\psi + \overline{\eta}(/\!\!\!p - /\!\!\!\!A - \kappa )\psi + \overline{\psi} (/\!\!\!p - /\!\!\!\!A - \kappa )\eta .
 \eqno(25) \]
The bilinear terms in eq. (15) can be written as
\[ \mathcal{L}_{2}  = \frac{1}{2}(A_\mu ,B_\mu )
\left( \begin{array}{cc}
a^{\mu\nu} & a^{\mu\nu} \\
a^{\mu\nu} & 0 \end{array} \right)
\left( \begin{array}{c}
A_\nu \\
B_\nu\end{array} \right)
 + (\overline{\psi}, \overline{\eta} )
 \left( \begin{array}{cc}
K& K \\
K & 0 \end{array} \right)
\left( \begin{array}{c}
\ \psi \\
\ \eta\end{array} \right)\eqno(26)
 \]
where $a^{\mu\nu} = \left(\partial^2 - m^2\right) \eta^{\mu\nu} - \partial^{\mu}\partial^{\nu}$ and $K = /\!\!\!p - \kappa$. Since

\[
\left( \begin{array}{cc}
a^{\mu\nu} & a^{\mu\nu} \\
a^{\mu\nu} & 0 \end{array} \right)^{-1} =
\left( \begin{array}{cc}
 0 & a_{\mu\nu}^{-1}  \\
a_{\mu\nu}^{-1} & -a^{-1}_{\mu\nu}     \end{array} \right)
\quad
\left( a_{\mu\nu}^{-1} = \frac{\eta_{\mu\nu} - \partial_\mu\partial_\nu/m^2}{\partial^2 - m^2}\right) \eqno(27)\]

\[
 \left( \begin{array}{cc}
K& K \\
K & 0 \end{array} \right)^{-1} =
\left( \begin{array}{cc}
0 & K^{-1} \\
K^{-1} & -K^{-1}
\end{array} \right) \quad \left( K^{-1} = 1/(/\!\!\!p - \kappa)\right)
\eqno(27) \]
we see that the propagators $<AA>$ and $<\psi \overline{\psi}>$ do not follow from $\mathcal{L}$.  As all vertices are at most linear in $B_{\nu}$, $\eta$ and $\overline{\eta}$, we see that there are no diagrams beyond one-loop order in the loop expansion for the effective action.

This approach to generating a unitary and renormalizable theory for massive vectors may be applicable to the Standard Model. The study of this case is presently in progress [33]. It may be necessary to consider such alternatives, both since the Higgs particle is proving to be difficult to detect, and since the radiative corrections to the Higgs potential appear to ``flatten'' it [34-36].

It has been noted [37] that in the first order form of the Einstein-Cartan action in $2 + 1$ dimensions, the loop expansion about a vanishing dreibein and spin connection terminates at one-loop order because the dreibein field enters the classical action only linearly.  The same situation occurs when considering the first-order Einstein-Hilbert action in $1 + 1$ dimensions [38].

\section*{Acknowledgements}
T.N. Sherry and J.W. Moffat had helpful comments.  Roger Macleod stimulated this research.

\end{document}